\documentclass[aps,reprint,amssymb]{revtex4-2}

\usepackage{graphicx}% Include figure files
\usepackage{dcolumn}% Align table columns on decimal point
\usepackage{bm}% bold math

\usepackage[utf8]{inputenc}
\usepackage[T1]{fontenc}
\usepackage{mathptmx}

\usepackage[version=4]{mhchem}%chemicals
\usepackage{hyperref}%links
\eprint

\begin{document}

\title{Magnetotransport in ferromagnetic \ce{Fe2Ge} semimetallic thin films}
\author{Andrew W. Forbes}
\email[Electronic mail: ]{59forbes@cua.edu}
\author{Niraj Bhattarai}
\author{Christopher Gassen}
\author{Raghad S. H. Saqat}
\author{Ian L. Pegg}
\author{John Philip}
\affiliation{The Vitreous State Laboratory, Washington, D.C. 20064, USA}
\affiliation{Department of Physics, The Catholic University of America, Washington D.C., 20064, USA}

\date{\today}

\begin{abstract}
Thin films of the ferromagnet \ce{Fe2Ge} were grown via molecular beam epitaxy, and their electrical and magneto-transport properties measured for the first time. X-ray diffraction and vibrating sample magnetometry measurements confirmed the crystalline ferromagnetic \ce{Fe2Ge} phase. The observed high temperature maximum in the longitudinal resistivity, as well as the observed suppression of electron-magnon scattering at low temperatures, point to the presence of strong spin polarization in this material. Measurements of the Hall resistivity, $\rho_{xy}$, show contributions from both the ordinary Hall effect and anomalous Hall effect, $\rho_{xy}^{AH}$, from which we determined the charge carrier concentration and mobility. Measurements also show a small negative magnetoresistance in both the longitudinal and transverse geometries. \ce{Fe2Ge} holds promise as a useful spintronic material, especially for its semiconductor compatibility.
\end{abstract}

\maketitle

\section{\label{sec:level1} Introduction}
The \ce{Fe-Ge} system contains an unusually high number of distinct crystallographic phases, with twelve phases of intermediate stoichiometry \cite{Khalaniya2019}. Many \ce{Fe-Ge} phases exhibit interesting magnetic properties, such as  cubic \ce{FeGe}, which can maintain helimagnetic order at room temperature \cite{Zhang2017}, and \ce{FeGe2} which can reportedly host spin density waves \cite{Tang2017}.

In this paper, we have used molecular beam epitaxy to grow the iron-rich hexagonal $\beta$ phase. The crystal structure and magnetism of bulk \ce{Fe2Ge} were extensively investigated in the 1960s \cite{Yasukochi1961,Katsuraki1964,Kanematsu1965,Kanematsu1965a,Adelson1965,Richardson1967}. Due to a search for suitable spintronic materials, a renewed interest in \ce{Fe2Ge} has emerged, with a few experimental studies concerning the thin film properties of the material, focused mainly on the ability of the (0001) crystal plane of Fe$_{1.7}$Ge to be grown epitaxially on the Ge(111) crystal plane \cite{Jaafar2008,Jaafar2010}. Recently, monolayer \ce{Fe2Ge} has been theoretically predicted to exhibit massive Weyl points due to nontrivial topology \cite{Liu2020}. However, unlike many other iron-germanium phases, to our knowledge there have been no studies on the electrical or magneto-transport properties of \ce{Fe2Ge}.

Care should be taken to distinguish the $\beta$ and $\eta$ phases of the \ce{Fe-Ge} system which are quite similar hexagonal structures.  The $\beta$ and $\eta$ phases are non-stoichiometric compounds that nonetheless maintain an ordered crystallographic structure. The $\beta$ structure is reported to be fully ordered at Fe$_{1.67}$Ge, but has iron vacancies at the d sites. As iron content is increased, these vacancies fill, but additional d-c site disorder appears.\cite{Kanematsu1964} As germanium content increases, additional a-d site disorder appears. The $\eta$ phase is exactly the size of two unit cells of the $\beta$ phase. However, the $\eta$ phase is not just case of additional long-term order (crystal supercell), but rather a totally distinct phase determined by a subtle order-disorder transition \cite{Kanematsu1965}. There are some differing reports on the homogeneity range. By all accounts the $\beta$ and $\eta$ phases are stable at high temperature, and metastable at room temperature. However, the eutectoid phase separation occurs slowly, and faster for iron-rich stoichiometries. Thus, the exact homogeneity range is not straightforward, and depends on the details of the experimental synthesis and measurement. In thin films, $\beta$ is reported to be homogeneous between \ce{Fe2Ge} to Fe$_{1.5}$Ge, while $\eta$ is homogeneous between Fe$_{1.4}$Ge and Fe$_{1.3}$Ge \cite{Jaafar2010}.

The $\beta$ phase is most often identified with the \ce{Ni2In} structure type ($B8_2$), with space group P6$_3$/mmc. However, based on slight corrections to the positions of the interstitial atoms, there is growing support for identifying the $\beta$ structure with space group P$\bar{3}$m1 \cite{Malaman1980,Jaafar2010}. The subtle breaking of sixfold point symmetry could also explain the existence of the observed in-plane uniaxial anisotropy \cite{Jaafar2010}, which was unable to be accounted for by strict hexagonal symmetry. This $\beta$-\ce{Fe2Ge} structure would then match the $\beta$-\ce{Fe2Si} structure, which has space group P$\bar{3}$m1 and is also is a slight deviation from the \ce{Ni2In} structure \cite{Errandonea2008}. Regardless, the $\beta$ phase is often still indexed with P6$_3$/mmc \cite{Merkel2019}. The minor differences between these structures are explored in detail in the supplemental material.

\section{Experimental Procedure}

Silicon wafers cut along the (100) direction were used as substrates. They were cleaned and etched with concentrated hydrofluoric acid to remove the native oxide on the surface. After being placed in a vacuum system, a base layer of MgO of thickness  5 nm was deposited, then a thin film of \ce{Fe2Ge} with thickness t $\approx$ 50~nm was deposited by molecular beam epitaxy. Pressure was measured by a Bayard-Alpert type hot ionization gauge, thickness was monitored by a quartz crystal microbalance, and deposition was controlled by a mechanical shutter. The films were annealed \emph{in situ} by a joule heater for two hours at 600~K. Finally, a protective MgO layer with thickness 5~nm was deposited to prevent oxidation.

The resulting thin films were analyzed by x-ray diffraction in the Bragg-Brentano geometry by Cu-K$_{\alpha}$ radiation ($\lambda=0.154059$~nm) in the double angular range 10$^\circ$$\leq2\theta\leq$90$^\circ$ (ARL X'TRA, Thermo Fischer Scientific, Waltham, Massachusetts). The resulting reflections were indexed to crystalline planes in the \ce{Fe2Ge}, and the details of the structure (dimensions of the unit cell, c-axis positions of the interstitial atoms, etc.) were determined using Rietveld refinement.

The magnetic moment of the films was determined by vibrating sample magnetrometry measured by a Physical Property Measurement System (Quantum Design, San Diego, California). The magnetic moment per unit cell, $m_{cell}$ (units $\mu_B/$cell), was calculated using the ratio of the macroscopic volume of the thin film, $V_{total}$ and the volume of the unit cell as determined by Rietveld refinement, $V_{cell}$.

\begin{equation}
m_{cell}=M_{total}\cdot \frac{V_{cell}}{V_{total}}
\end{equation}

Then, circuits were built to analyze the longitudinal resistivity ($\rho_{xx}\equiv E_x/J_x$) and transverse resistivity ($\rho_{xy}\equiv -E_y/J_x$) using gold wire bonded to the film by indium metal contacts. The longitudinal resistivity was measured using a standard four-probe method, with a geometric correction factor \cite{Smits1958},

\begin{equation}
\rho_{xx}=\rho_{s}\cdot t=\frac{V_x}{I_x}C_0 \cdot t
\end{equation}

where $\rho_{s}$ is the sheet resistance, t is the thickness of the film, and $C_0$ is a correction factor determined by the geometry. The transverse resistivity was measured by another portion of thin film deposited simultaneously using a shadow mask to create a Hall bar shape. The resulting circuits were placed in a physical property measurement system (PPMS) with the horizontal rotator AC transport option. A transverse voltage offset due to imperfect geometry was accounted for by the addition of a constant, $C_1$ under the assumption that $\rho_{xy} = 0$ at zero applied magnetic field, after accounting for any possible hysteresis.

\begin{equation}
\rho_{xy}=\frac{V_y}{I_x}t+C_1
\end{equation}

\section{Results and Discussion}

\subsection{Structural Properties}

\begin{figure}
\includegraphics[scale=0.8]{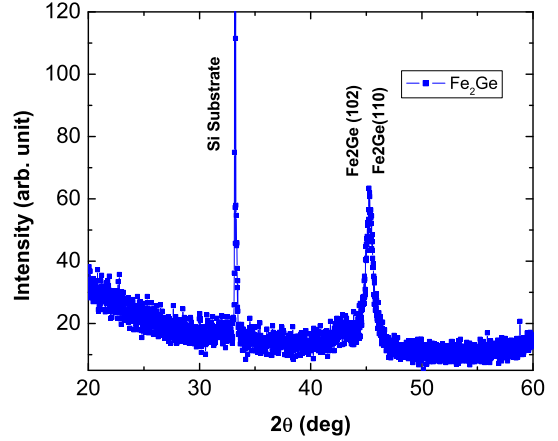}
\caption{X-ray diffraction measurements show the main (102) and (110) peaks of \ce{Fe2Ge} near a 2$\theta$ value of $45^\circ$.}
\label{fig:XRD}
\end{figure}

As shown in figure \ref{fig:XRD}, the x-ray diffraction pattern obtained from \ce{Fe2Ge} is explained by a single phase with lattice constants $a=3.99$~\AA, $c=4.89$~\AA. This is slightly smaller than the previously experimentally reported P6$_{3}$/mmc structure $a=4.06$~\AA, $c=5.04$~\AA \cite{Adelson1965}, but quite comparable to the theoretical P$\bar{3}$m1 structure, which is predicted $a=4.01$~\AA, $c=4.89$~\AA \cite{Jain2013}.

The positions of the interstitial atoms were consistent with the P6$_{3}$/mmc structure, though P$\bar{3}$m1 was not ruled out. The quality of the fit was determined by the standard parameters such as the weighted profile R-factor, $R_{wp}=16.82\%$, which quantifies the difference between the observed intensities, $y_{o,i}$ and the calculated intensities based on the theoretical structure, $y_{c,i}$ according to a weighting function, $w_i$.

\begin{equation}
R_{wp}=\sqrt{\frac{\Sigma_i w_i (y_{o,i} -y_{c,i})^2}{\Sigma_i w_i y_{o,i}^2}} x 100\%
\end{equation}
\begin{equation}
w_i=\frac{1}{\langle(y_{o,i}-\langle y_{o,i}\rangle )^2\rangle}
\end{equation}

The expected R-factor, $R_{exp}$, and the $\chi^2$, are defined by

\begin{equation}
R_{exp}=\frac{N}{\Sigma_i w_i (y_{o,i})^2}
\end{equation}
\begin{equation}
\chi^2=(\frac{R_{wp}}{R_{exp}})^2
\end{equation}

\subsection{Magnetic Properties}

\begin{figure}
\includegraphics[scale=0.8]{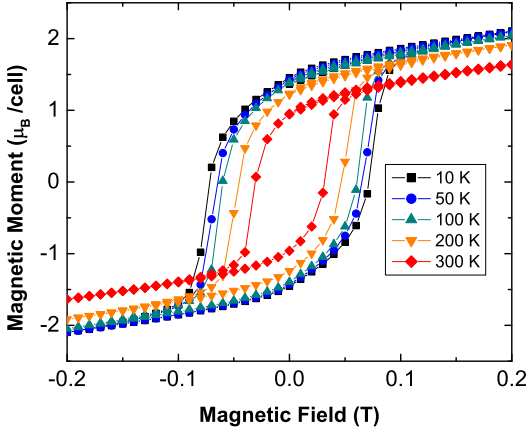}
\caption{Magnetic moment versus magnetic field at different temperatures shows ferromagnetic character across the whole temperature range, with monotonically increased coercivity and saturated moment at lower temperatures.}
\label{fig:MvH}
\end{figure}

\begin{figure}
\includegraphics[scale=0.8]{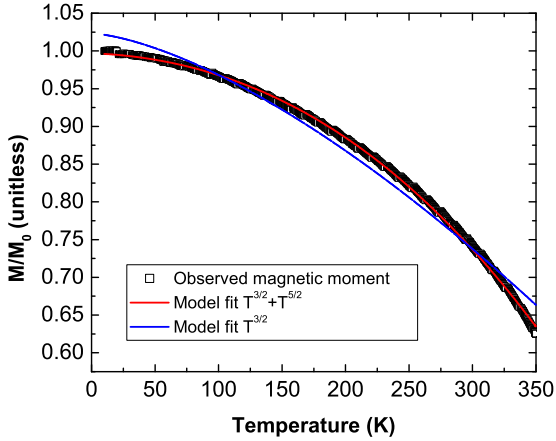}
\caption{The observed magnetic moment versus temperature curve under an applied magnetic field of 200~Oe shows a consistent magnetic phase with a moment that decreases with temperature. The fit predicts a Curie Temperature $T_c=565$~K which is slightly higher than the previously observed $T_c=500$~K in the bulk system \cite{Kanematsu1965}. The fit according to equation \ref{Eq:MvTsimple} (red solid line) clearly does not explain the data nearly as well as well as the fit of the form of equation \ref{Eq:MvTcomplex} (blue dashed line).}
\label{fig:MvT}
\end{figure}

The magnetization versus applied magnetic field plot shown in figure \ref{fig:MvH} shows a magnetic hysteresis behavior that indicates the material is ferromagnetic, with a stronger magnetic moment at low temperatures. As shown in the magnetization versus temperature plot in figure \ref{fig:MvT}, \ce{Fe2Ge} displayed a monotonic, smoothly decreasing magnetization with increasing temperature. This is indicative of a single system with no magnetic transitions in the temperature range 10 $\leq$ T $\leq$ 350~K. As shown by the line, this behavior was fit to the Bloch $T^{3/2}$ law, which is of the form

\begin{equation}
M(T)=M_0(1-B T^{3/2})
\label{Eq:MvTsimple}
\end{equation}

The best-fit parameters give $B=5.37 \times~10^{-5}$~$K^{-3/2}$, with an adjusted $R^2$ of 0.999702.

However, looking at figure \ref{fig:MvT} this leading term of the spin-wave model (shown in the dashed blue line) is not an ideal match for the observed magnetization dependence. To account for this, we consider the $T^{3/2}$ and $T^{5/2}$ terms from the Bloch-Dyson theory, also considering the possibility of an energy gap \cite{Keffer1966,Shu2021},

\begin{equation}
\begin{aligned}
M(T)=M_0(1-a_{3/2}[F(\frac{3}{2},t_H)/\zeta(\frac{3}{2})]T^{3/2} \\
-a_{5/2}[F(\frac{5}{2},t_H)/\zeta(\frac{5}{2})]T^{5/2})
\end{aligned}
\label{eq:MvT}
\end{equation}

where $\zeta(s)$ is the Riemann zeta function, $a_{3/2}$ and $a_{5/2}$ are constants, and $F(s,t_H)$ is the Bose-Einstein integral

\begin{equation}
F(s,t_H)=\sum_{p=1}^{\infty} p^{-s}e^{-p/t_H}\approx e^{-1/t_H}
\end{equation}

This exponential approximation is often used for the Bose-Einstein integral \cite{Niira1960,Forbes2019}. Here $t_H$ is based on a spin wave energy gap, $\Delta$, of the form $t_H=k_B T/\Delta$. Rewriting equation \ref{eq:MvT} this way gives:

\begin{equation}
\begin{aligned}
M(T)=M_0(1-a_{3/2}[e^{-\Delta/k_B T}/\zeta(\frac{3}{2})]T^{3/2} \\
-a_{5/2}[e^{-\Delta/k_B T}/\zeta(\frac{5}{2})]T^{5/2})
\end{aligned}
\label{eq:MvTgap}
\end{equation}

Fitting the curve with the function in equation \ref{eq:MvTgap} (not shown), we get a negligibly small negative spin wave excitation gap. Since a negative spin wave excitation gap is non-physical, we instead conclude the gap is not present, $\Delta \approx 0$. This contrasts with \ce{Fe2Si}, which has a positive gap \cite{Forbes2019}.

Since there is zero gap, in the case of \ce{Fe2Ge} we can simplify the unapproximated equation \ref{eq:MvT} using the fact that $F(\frac{3}{2},t_H)=Li_{3/2}(e^{-1/t_H})$ and $Li_x(1) = \zeta(x)$ leading directly to:

\begin{equation}
M(T)=M_0(1-a_{3/2}T^{3/2} -a_{5/2}T^{5/2})
\label{Eq:MvTcomplex}
\end{equation}

As shown in the solid red line, this equation was fitted to magnetization curve in figure \ref{fig:MvT}. The resulting constants from this fit ($a_{3/2}=1.83 \times~10^{-5}$ and $a_{5/3}=1.06 \times~10^{-7}$) are similar to other metals \cite{Keffer1966}. The adjusted $R^2$ is an improved 0.999985.

The constants are determined by the dispersion coefficient, D, for example by:

\begin{equation}
a_{3/2}=\zeta(\frac{3}{2})\frac{g\mu_B}{M_0}(\frac{k_B}{4\pi D})^{3/2}
\end{equation}

Finding $M_0$ based on the dimensions of my sample, $D/k_B$=$4.54 \times~10^{-13}$, similar to pure iron at $3.32 \times~10^{-13}$ and nickel at $4.66 \times 10^{-13}$ \cite{Keffer1966}.

\subsection{Longitudinal resistivity}

\begin{figure}
\includegraphics[scale=0.8]{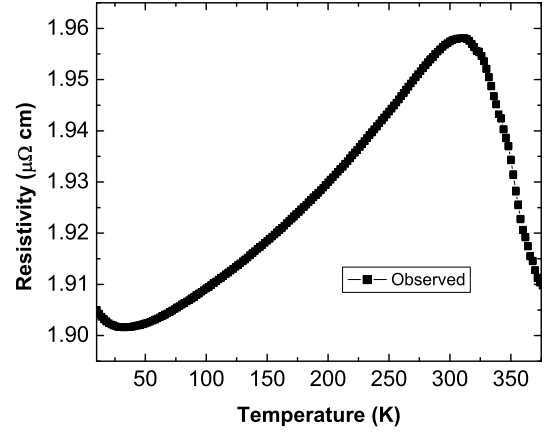}
\caption{The Resistivity is metallic below 300~K, but begins decreasing above 312~K.}
\label{fig:RxxvsT}
\end{figure}

The longitudinal resistivity, $\rho_{xx}$, was measured from 10 to 375~K (see figure \ref{fig:RxxvsT}). It shows metallic behavior up to around 300~K, where it undergoes a drop in resistivity. This drop in resistivity could be due to a change in crystal structure, though there has been no other evidence of such a structural transition.  Another possibility is shown in its close resemblance to the behavior seen in cobalt-based half-metallic ferromagnets \ce{Co2VAl} and \ce{Co2CrGa}, which show similar drops in resistivity above a metallic regime but below the Curie temperature \cite{Kourov2013,Kourov2016}. The curve was fit in several parts accounting for which interactions are important in the various temperature ranges.

First, we consider the low temperature part of the resistivity. A common type of resistivity fit for such half-metallic materials follows the square of the magnetization, which leads to the electron-magnon resistivity having a $T^2$ dependence \cite{Kourov2016}.

\begin{equation}
\rho_{e-m}=cM^2=\rho_1+\rho_2(T/T_c)^2
\end{equation}

This $T^2$ dependence fits observations reasonably well at intermediate and high temperatures, but a detailed low temperature fit reveals the presence of a low-temperature energy gap, $\Delta$, which suppresses the effect of electron-magnon scattering due to the inaccessibility of down-spin electron states with energies less than the typical magnon energies for the conduction electrons to be scattered to \cite{Barry1998}. Such a gap could be attributable to a gap in the magnon excitation spectra, but this possibility is ruled out since (as discussed in the magnetization section) no such magnon excitation gap was observed in this sample. Therefore, the reduced electron-magnon scattering at low temperatures can only be attributed to a gap in the electronic spectrum, a signal of a half-metallic band structure.

\begin{figure}
\includegraphics[scale=0.8]{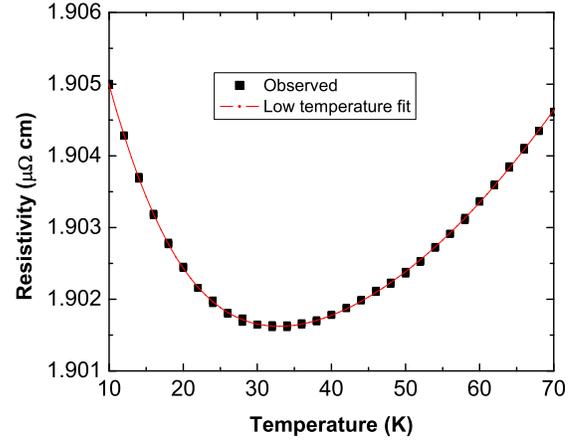}
\caption{The low temperature fit observed from 10 to 70~K, fit according to equation \ref{Eq:lowt}.}
\label{fig:RxxvsTLow}
\end{figure}

The resistivity measurements in the low temperature regime (T $\leq$ 70~K) can be fitted with the following expression yielding the constants in Table \ref{tab:Rxxfit}, as shown in figure \ref{fig:RxxvsTLow}:

\begin{eqnarray}
\rho_{xx}=\rho_0-\gamma_{e-e}T^{1/2}+\beta_{e-m}T^2e^{-\Delta/T}\nonumber \\
+\alpha_{e-p}\int_0^{\theta_D/T}\frac{x^5}{(e^x-1)(1-e^{-x})}dx
\label{Eq:lowt}
\end{eqnarray}

The medium temperature range was fit with a metallic behavior \cite{Forbes2019}. Here, a better fit was obtained with a simple T dependence rather than the complete Bloch-Grueisen integral, which is consistent with the fact that this integral reduces to a linear dependence on temperature in the high temperature limit $T \gg \Theta_D$. Also in this range the energy gap in the electron-magnon contribution is no longer relevant.

\begin{figure}
\includegraphics[scale=0.8]{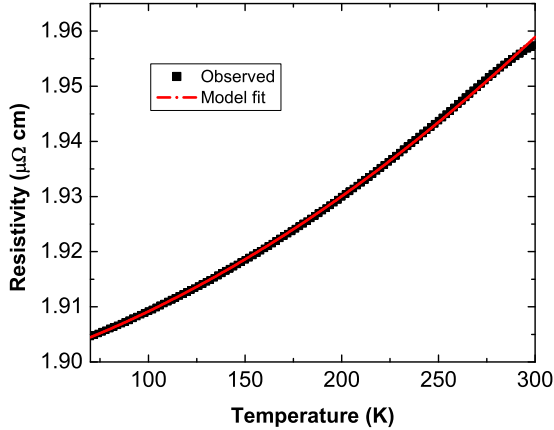}
\caption{The medium temperature fit observed from 70 to 300~K, fit according to equation \ref{Eq:medt}.}
\label{fig:RxxvsTMed}
\end{figure}

The medium-temperature metallic regime (70 $\leq$ T $\leq$ 300~K) can be fit with the following equation yielding the constants in Table \ref{tab:Rxxfit}, as shown in figure \ref{fig:RxxvsTMed}:

\begin{equation}
\rho_{xx}=\rho_0+\beta_{e-m}T^2+\alpha_{e-p} T
\label{Eq:medt}
\end{equation}

At around 300~K, this model begins to diverge from the observed data, due to the appearance of semimetallic behavior. The high-temperature data shows semiconductor behavior which is better explained by a simplified Steinhart-Hart equation \cite{Steinhart1968}.

\begin{figure}
\includegraphics[scale=0.8]{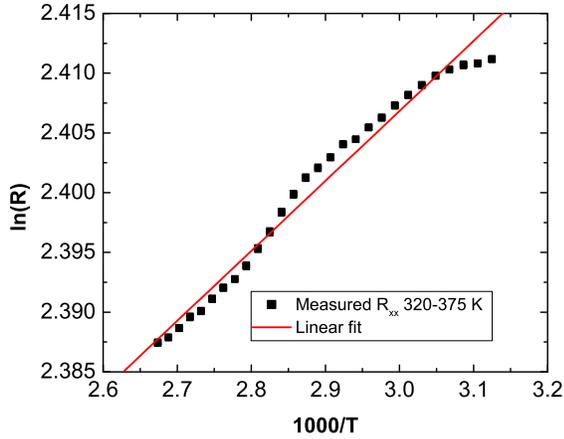}
\caption{The high temperature fit observed from 320 to 375~K, fit according to equation \ref{Eq:hight}. This shows semimetallic behavior consistent with a narrow band gap of 5.0~meV}
\label{fig:RxxvsTHigh}
\end{figure}

Semiconducting regime (320 $\leq$ T $\leq$ 375~K, figure \ref{fig:RxxvsTHigh}):

\begin{equation}
ln(R)=ln(R_0)+\frac{E_a}{k_BT}
\label{Eq:hight}
\end{equation}

Fitting to this equation leads to a semiconducting energy gap of 5.0~meV, as shown in figure \ref{fig:RxxvsTHigh}.

\begin{table*}
\begin{ruledtabular}
\begin{tabular}{cccccccc}
\textrm{Domain}&
\textrm{$1-R^2$}&
\textrm{$\rho_0$ ($\mu\Omega$cm)}&
\textrm{$\gamma_{e-e}$ (n$\Omega$cm/$K^{1/2}$)}&
\textrm{$\beta_{e-m}$(p$\Omega$cm/$K^2$)}&
\textrm{$\Delta$(K)}&
\textrm{$\alpha_{e-p}$ (n$\Omega$cm)}&
\textrm{$\Theta_D$(K)}\\
\colrule
$T\leq$70~K&$6.60 \times 10^{-11}$&1.91&2.48&1.61&110&93.8&126 \\
$70~K \leq T \leq 300~K$&$1.90 \times 10^{-8}$&1.90&&0.400&&0.0887&\\
\end{tabular}
\end{ruledtabular}
\caption{The fitting constants of equation \ref{Eq:lowt} as shown in figure \ref{fig:RxxvsTLow}, and the fitting constants of equation \ref{Eq:medt} as shown in figure \ref{fig:RxxvsTMed}.}
\label{tab:Rxxfit}
\end{table*}

\subsection{Hall effect}

\begin{figure}
\includegraphics[scale=0.8]{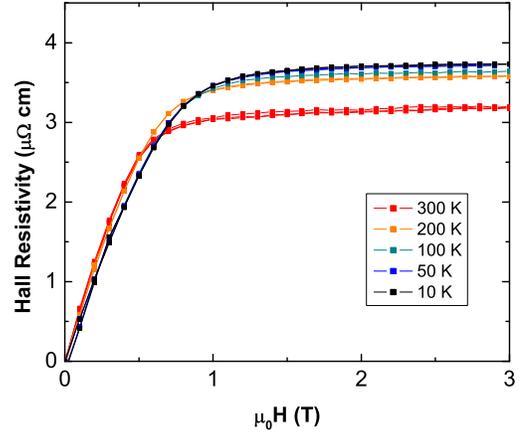}
\caption{ Hall resistivity versus Field measurements at various temperatures shows monotonically higher resistivity at lower temperatures.}
\label{fig:RxyvsH}
\end{figure}

The Hall resistivity, $\rho_{xy}$, was measured at different temperatures from 10 to 300~K (see figure \ref{fig:RxyvsH}). By performing a linear fit on the Hall resistivity well above saturation (1.5 to 3~T), we can infer the concentration of charge carriers in the material according to the single charge carrier model from the slope of the Hall resistivity plotted against magnetic field.

\begin{equation}
\rho_{xy}=\rho_{AH}+R_0B
\end{equation}

The mobility is inferred by combining the longitudinal resistivity and the Hall constant:

\begin{equation}
\mu_H=R_0/\rho_{xx}
\end{equation}

The Hall constants, $R_0$, and Hall mobilities, $\mu_H$, are summarized for the one-band model in Table \ref{tab:Hall}. At 10~K, $R_0=2.86 \times 10^{-8}$~$\Omega$ cm/T, or a charge carrier density of $n=2.18 \times 10^{22}$~cm$^{-3}$. These values are similar to the values previously obtained for the related compound \ce{Fe2Si} \cite{Forbes2019}.

\begin{table}
\begin{ruledtabular}
\begin{tabular}{ccccc}
\textrm{T (K)}&
\textrm{$R_0$ (n$\Omega$cm/T)}&
\textrm{n} (cm$^{-3}$)&
\textrm{$\rho_{xx}$($\mu\Omega$ cm)}&
\textrm{$\mu_H$ ($cm^2/Vs)$)}\\
\colrule
10&36.5&$1.71 \times 10^{22}$&1.90&192\\
50&37.6&$1.66 \times 10^{22}$&1.90&198\\
100&34.4&$1.75 \times 10^{22}$&1.91&187\\
200&33.8&$1.85 \times 10^{22}$&1.93&175\\
300&48.2&$1.30 \times 10^{22}$&1.96&246\\
\end{tabular}
\end{ruledtabular}
\caption{The carrier concentration, resistivity and mobility as calculated for \ce{Fe2Ge} measured at various temperatures.}
\label{tab:Hall}
\end{table}

\subsection{Magnetoresistance}

\begin{figure}
\includegraphics[scale=0.8]{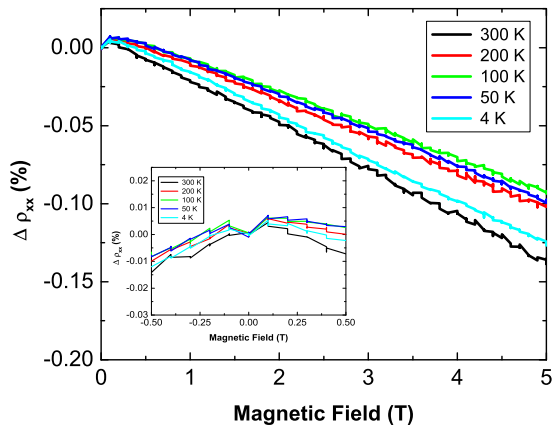}
\caption{The observed longitudinal magnetoresistivity is very similar to the transverse geometry. The inset shows the small drop in resistivity near zero magnetic field.}
\label{fig:longM}
\end{figure}

\begin{figure}
\includegraphics[scale=0.8]{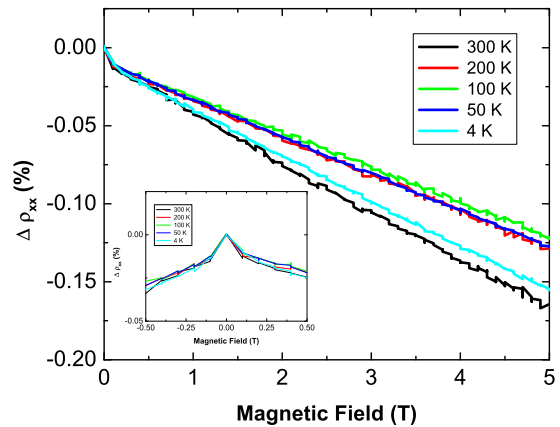}
\caption{The transverse magnetoresistivity has a very small dependence on field up to 5~Tesla,  modifying the resistivity by less than 0.2\%. The inset shows a small bump in resistivity around zero magnetic field.}
\label{fig:transM}
\end{figure}

Magnetoresistance ($\rho_{MR}$) measurements were carried out by the four-probe method, using both current parallel to the applied magnetic field (figure \ref{fig:longM}) and current transverse to the applied magnetic field (figure \ref{fig:transM}). The results are expressed as a percentage:

\begin{equation}
\rho_{MR}=(\rho_{xx}(B)/\rho_{xx}(0)-1)\times 100\%
\end{equation}

In both geometries, we observed a small negative magnetoresistance which was approximately linear in field up to 5~T. The negative magnetoresistance is attributable to the suppression of s-d interband scattering in magnetic fields, and is commonly seen in other 3d ferromagnets \cite{Raquet2001}. The low-field behavior of the two geometries is slightly different. The transverse magnetoresistance shows a larger negative slope at low fields, likely due to reduced scattering at the magnetic domain walls once the magnet is saturated. The longitudinal magnetoresistance shows a small but clear dip at zero field. This is attributable to an out-of-plane anisotropy induced because the magnetic field was not perfectly aligned to the current direction \cite{Eden2019}.

\section{Conclusion}

In this paper, we have examined the crystal properties, magnetic properties, and magneto-transport properties of the semimetal \ce{Fe2Ge} films in detail. The longitudinal resistivity showed a maximum at 312~K, as well as a reduced electron-magnon scattering at low temperatures. This behavior was explained in terms of a spin-flip excitation gap due to strong spin polarization at the fermi level. The Hall resistivity was used to measure the carrier concentration and carrier mobility, confirming the metallic nature of the material at and below room temperature. The material displayed a small magnetoresistance, consistent with similar transition metal ferromagnets. \ce{Fe2Ge} would be a useful platform for use in applications such as the sources for spintronic devices, which have seen considerable optimization in the manipulation of spins in the channel, but where efficient spin injection remains challenging \cite{Chuang2014,InglaAynes2021}.

\bibliographystyle{apsrev4-2}
\bibliography{Main_text_file}

\end{document}